\documentclass[conference]{IEEEtran}
\IEEEoverridecommandlockouts
\usepackage{cite}
\usepackage{amsmath,amssymb,amsfonts}
\usepackage{bm}
\usepackage{subcaption}
\usepackage[linesnumbered,ruled,noend]{algorithm2e}
\usepackage[export]{adjustbox}
\usepackage{threeparttable} 
\usepackage{xcolor}

\begin{document}
\twocolumn
\newtheorem{definition}{Definition}[section]

\title{AutoFlows++: Hierarchical Message Flow Mining for System on Chip Designs 
}

\author{\IEEEauthorblockN{Bardia Nadimi, Hao Zheng}
\IEEEauthorblockA{\textit{Bellini College of AI, Cybersecurity and Computing} \\
\textit{University of South Florida}\\
Tampa, United States \\
\{bnadimi, haozheng\}@usf.edu}
}

\maketitle
\vspace*{-15pt}

\begin{abstract}

Understanding communication behavior in modern system-on-chip (SoC) designs is critical for functional verification, performance analysis, and post-silicon debugging. 
Communication traces capture message exchanges among system components and provide valuable insights into system behavior. 
However, deriving concise communication specifications from such traces remains challenging due to interleaved instances of communication flows, and ambiguous causal relationships among messages.
Existing mining approaches often struggle with scalability and ambiguity when traces contain complex interleaving of message patterns across multiple components. 
These conditions often lead to an explosion in the number of candidate flows and inaccurate extraction of communication behaviors.
This paper presents AutoFlows++, a design-architecture-guided hierarchical framework for mining message flows from communication traces of complex SoC designs. 
AutoFlows++ operates in two stages: local mining followed by global mining. 
In the local mining stage, simple communication patterns are extracted from traces observed at individual communication interfaces between components. 
In the global mining stage, these local patterns are composed to identify higher-level message flows that characterize communication behavior across multiple components.
Experimental results on both synthetic traces and traces generated from SoC models in GEM5 demonstrate that AutoFlows++ significantly improves flow extraction accuracy compared with prior approaches, highlighting its effectiveness for practical SoC validation tasks.
\end{abstract}

\begin{IEEEkeywords}
system-on-chip, specification mining, model inference, message path mining, interface slicing 
\end{IEEEkeywords}

\section{Introduction}

\begin{figure}[tb]
\begin{center}
\begin{tabular}{p{1.4in}p{1.2in}}
\begin{minipage}{1.1in}
\centering
\includegraphics[width=1.1in]{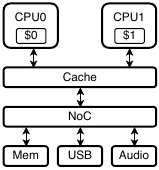}
\end{minipage}
& 
\begin{minipage}{1.2in}
\centering
\includegraphics[width=1.2in]{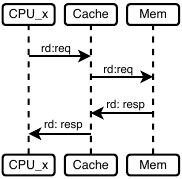}
\end{minipage}
\\
\centering (a) & \centering (b)
\end{tabular}
\caption{ (a) Simplified SoC architecture example, (b) Message sequence diagram for CPU downstream read flows, parameterized by $x=0$ or $1$. Only CPU, Cache, and Mem communications are shown for clarity.}
\vspace*{-10pt}
\label{fig:soc-ex}
\end{center}
\vspace*{-15pt}
\end{figure}

Modern system-on-chip (SoC) designs integrate a large number of heterogeneous functional blocks connected by  on-chip interconnect.
These blocks coordinate via sophisticated communication protocols by exchanging messages to realize various system functions. 
These system-level communication protocols are referred to as \emph{message flows}.
The correctness and performance of these designs critically depend on the validity of their communication behaviors. 
However, these blocks typically operate concurrently, leading to simultaneous or interleaved system transactions, which are a major source of various errors at the design time or runtime.
Fig.~\ref{fig:soc-ex} shows a simplified example of a SoC architecture integrating typical components such as CPUs, caches, and Network-on-Chip (NoC) interconnects.

Well-defined, accurate, and comprehensive specifications on system-level behaviors are essential for various activities including design, verification, validation and debug, etc.
Ideally, such specifications should be ready at the start of a SoC design, and updated continuously as the design evolves.
However, in reality, such specifications are usually non-existent,  ambiguous,  incomplete, or even contain errors.
It is also common that specifications become out-of-dated and disconnected from the implementations as the design moves across stages of its life cycle.
This phenomenon has been observed in software~\cite{Zhong:mine-api:2013} and complex hardware system designs~\cite{ray:2016:iccad,chen:2017:dnt}.
Finally, manually writing and maintaining such system level specifications is a tedious and unscalable task.
Consequently, the lack of such well-defined and comprehensive specifications renders the systematic and rigorous analysis and validation of system designs infeasible.

Communication traces of an SoC design can be obtained when such a design is validated via simulation or silicon execution.
These traces capture message exchanges between components of the design and are the result of message flows exercised during system execution. 
Given a large system integrating many design blocks, system communication traces from full-system validation are often very long with a large number of message flows executed concurrently.   
As a result, directly gaining insights into system communication behaviors from these traces is unwieldy.
The goal of this work is to analyze such traces, and automatically extract the embedded message flows.
The extracted message flows provide users an abstract view of system-wide protocols implemented in the target system and can be used for various validation and debugging tasks.
However, extracting valid and meaningful message flows from the communication traces is highly challenging as the communication traces are the results of message flows execute concurrently. 
Messages from different message flows are arbitrarily interleaved, obscuring their dependencies.
The same messages may be involved in different message flows, making it difficult to determine which message flows an observed message actually belongs to. 

Previous pattern mining approaches~\cite{Yang:2006,Liu:2013,Mrowca:2019:LTS:3316781.3317847,natasa2020} are inadequate for handling the SoC communication traces considered in this work. 
They extract patterns or models from a trace based on strong temporal dependencies among events in the trace. 
In the case of communication traces resulting from concurrent executions of message flows, messages showing strong temporal dependencies in traces may not be correlated according to the true dependencies in the ground truth specifications. 
Consequently, the existing mining methods often produce numerous patterns, many of which are meaningless or invalid. Moreover, these methods may fail to extract a model encompassing all valid patterns, leading to large, incomprehensible, or even misleading extracted models.

In this paper, we present AutoFlows++, a hierarchical framework guided by design architecture for mining message flows from communication traces of complex SoC designs. 
Since an SoC design can be viewed as a network of components, AutoFlows++ operates in two stages: local mining followed by global mining.
In the local mining stage, simple communication patterns are extracted from traces observed at individual communication interfaces between components. 
The rationale for local mining is that these simple communication patterns can be mined with higher accuracy, since the communication scenarios between individual components are much simpler compared with system-level communication scenarios involving many components communicating concurrently.
In the global mining stage, system-level message flows that characterize communication behavior across multiple components are identified from the system communication traces by assembling the local patterns mined in the local mining stage.
This staged hierarchical mining framework makes mining sophisticated system-level message flows from very long traces with complex structures more tractable and accurate.
This paper makes the following \textbf{contributions}:
\begin{itemize}
    \item We propose a hierarchical mining framework guided by design architecture that automatically and efficiently infers accurate message flows from highly concurrent communication traces of complex system designs. 
    To the best of our knowledge, this is the first approach that leverages architectural structure to guide hierarchical mining of communication flows.

    \item We introduce a \emph{path energy based ranking method} that leverages locally mined communication patterns to prioritize candidate message flows, thereby significantly reducing the space of candidate communication models that must be explored.

    \item We develop a model evaluation method based on positional indexing that substantially improves the accuracy of the mined communication models.

    \item We conduct a comprehensive experimental evaluation using diverse synthetic traces as well as traces generated by running real software applications on realistic SoC models implemented in GEM5~\cite{gem5}. 
    The results demonstrate that the proposed framework can infer message flows with very high accuracy.
\end{itemize}
At a high level, AutoFlows++ reduces ambiguity by first isolating local interactions and then reconstructing global flows under positional constraints.


The remainder of this paper is organized as follows. 
Section~\ref{sec:relatedWorks_new} reviews related work. 
Section~\ref{sec:preliminaries} introduces the preliminaries and problem formulation. 
Section~\ref{sec:method} describes the AutoFlows++ framework. 
Section~\ref{sec:experimentalResults} presents the experimental evaluation of the proposed approach. 
Section~\ref{subsec:usecases} describes some practical use cases of AutoFlows++.
Finally, Section~\ref{sec:conclusion} concludes the paper and section~\ref{sec:future-work} outlines directions for future work.

\section{Related Works}
\label{sec:relatedWorks_new}

Specification mining from execution traces has been widely studied in software engineering, hardware verification, and system validation. 
In this section, we review prior work most relevant to AutoFlows++, including: (A) trace-based specification mining, (B) hardware assertion and protocol mining, (C) model synthesis from communication traces, and (D) concurrency-aware trace analysis. 
We then summarize the limitations of prior work and position AutoFlows++ with respect to these approaches.

\subsection{Trace-Based Specification Mining}

Early work on specification mining focused on extracting behavioral models from execution logs. 
Synoptic infers models of concurrent systems by mining temporal invariants and constructing finite-state abstractions consistent with observed traces \cite{synoptic:mine-fsm:2011}. 
Perracotta mines temporal API rules from software execution traces by discovering recurrent ordering constraints among events \cite{Perracotta}. 
BaySpec learns temporal specifications in the form of LTL formulas from Bayesian networks trained on execution traces \cite{BaySpec}. 
These techniques demonstrate that useful specifications can be inferred automatically from observed executions; however, they primarily target software traces and generally rely on temporal dependencies observed at the whole-trace level.

Beyond these approaches, process mining techniques provide a broader framework for extracting structured behavioral models from event logs. 
For example, van der Aalst’s process mining framework \cite{van2016process} and fuzzy mining \cite{gunther2007fuzzy} aim to derive high-level process models from noisy and incomplete logs. 
While effective in business-process domains, these methods typically assume well-defined event boundaries and struggle with highly interleaved traces, limiting their applicability to SoC communication traces.

Several later efforts have explored automata- or trace-based specification inference in domains with more structured and interleaved traces. 
Recent work on smart-contract specification mining combines trace slicing with abstraction refinement to recover automata from interleaved transaction histories \cite{SmartContracts}. 
Similarly, recent studies on trace-based debugging and analysis in distributed systems highlight the importance of structured trace representations to handle complex interleavings \cite{guo2020gmta}. 
While such methods share the objective of recovering high-level behavioral models from traces, they rely on domain-specific slicing keys and state abstractions that are not directly available for SoC communication traces.

\subsection{Hardware Assertion and Protocol Mining}

Within hardware verification, a substantial body of work has focused on mining assertions or temporal properties from traces and simulation artifacts. 
Existing approaches extract assertions from gate-level designs \cite{Li:2010:dac}, RTL models \cite{Hertz:2013:tcad,Danese:2015:vlsi-soc,Danese:2015:date,Danese:2017:dac,Chang:2010:aspdac}, and transaction-level models \cite{Liu:2013}. 
These techniques are valuable for discovering signal-level or protocol-level properties, but they are generally aimed at assertion generation rather than reconstructing complete message flows spanning multiple interacting components.

More recent work has continued to improve assertion mining efficiency and temporal expressiveness. 
For example, ARTmine uses association-rule mining with temporal behavior to generate compact assertion sets with improved coverage and lower verification cost \cite{iman2024artmine}. 
Additionally, recent efforts have explored combining static and dynamic analysis to improve assertion quality and scalability, further highlighting the importance of structured trace interpretation in hardware verification. 
Nevertheless, assertion-mining approaches typically focus on local temporal relations and property extraction, whereas AutoFlows++ targets explicit recovery of system-level communication flows from highly concurrent message traces.

\subsection{Model Synthesis from Communication Traces}

A line of work closely related to ours aims to synthesize formal models directly from execution traces. 
Classical automata-inference and model-synthesis approaches include SAT-based deterministic finite automata inference \cite{Heule:2013,Lang:1998:EDSM}, extended finite-state machine induction \cite{Ulyantsev:2011}, and trace-to-model style synthesis methods that aim to construct abstract behavioral models from observed traces \cite{Walkinshaw:2016}. 
In the SoC domain, model synthesis for communication traces formulates the problem as a constraint-solving task to infer finite-state models consistent with communication traces \cite{modelSynthesis}. 
AutoModel extends this direction by using acceptance ratio to guide model construction rather than relying purely on model size \cite{OurTCAD}.

The original AutoFlows framework moved beyond FSA composition and proposed directly inferring message flows from SoC communication traces using a causality graph, acceptance-ratio--guided refinement, and essential-causality optimizations \cite{AutoFlows}. 
Compared with AutoModel, this significantly improved the ability to capture longer and more semantically meaningful message sequences. 
However, AutoFlows still performs inference at the global-trace level and relies on heuristic evaluation and iterative refinement, making it vulnerable to ambiguity under heavy interleaving and costly on very long traces.

A related earlier study also examined message-flow mining from SoC traces in a more restricted setting, but focused primarily on sequential patterns rather than explicitly addressing highly concurrent interleaving \cite{Ahmed:mine-msg-flow:2021}. 
In addition, SEVNoC emphasizes validation of communication-fabric security properties in NoC-based SoCs, highlighting the importance of system-level communication analysis for verification \cite{Noc_fabrics}, although it does not address automatic flow mining from interleaved communication traces.

\subsection{Concurrency-Aware and Trace-Structuring Methods}

Interleaving and ambiguity are central challenges in trace analysis for complex systems. 
Foundational work in distributed systems has shown that temporal ordering does not necessarily imply causality, motivating the need for more structured reasoning over events \cite{lamport1978time,mattern1989virtual}. 
In domains outside hardware, recent work has shown that structured representations, slicing, and sequential-pattern analysis can substantially improve trace reasoning under complex interactions. 
For example, TraceContrast uses structured trace representations and contrast sequential pattern mining to localize root causes in microservice systems \cite{zhang2024tracecontrast}. 
Similarly, large-scale tracing infrastructures such as Dapper highlight the challenges of reconstructing causality from distributed traces \cite{sigelman2010dapper}. 

Although these approaches are not designed for specification inference, they reinforce the broader observation that preserving structural and positional relationships is essential when traces are long, heterogeneous, and highly concurrent. 
This observation is especially relevant for SoC communication traces, where messages from different flow instances are interleaved arbitrarily and where purely temporal correlations may not correspond to true protocol dependencies. 
AutoFlows++ addresses this challenge directly through hierarchical interface slicing, energy-guided path ranking, and position-aware evaluation.

Recent advances in large language models (LLMs) and machine learning techniques have also explored automation in hardware design, specification generation, and verification, including HDL code synthesis from high-level descriptions \cite{MultiExpertLLM, PyraNet, CodeV, verimind}, specification generation \cite{LLM_Spec_1, LLM_Spec_2, MLM_spec_gen}, and verification \cite{tbntb, LLM_verification_2}.

\subsection{Summary and Positioning}

Prior work has demonstrated the feasibility of mining specifications, assertions, and behavioral models from execution traces. 
However, existing approaches still lack several capabilities required for modern SoC communication traces: (1) architectural decomposition that reduces ambiguity before global inference, (2) explicit position-aware reasoning for resolving competing flow instances, and (3) a unified optimization objective for prioritizing candidate flows. 
AutoFlows++ addresses these limitations by integrating interface slicing, confidence-guided local validation, path-energy--based global ranking, and positional indexing into a single hierarchical framework. 
In this sense, AutoFlows++ extends AutoFlows from a global heuristic refinement approach into a hierarchical, position-aware, and optimization-driven framework for mining message flows from highly concurrent SoC traces.

To summarize the key differences between AutoFlows and the proposed AutoFlows++ framework, Table~\ref{tab:comparison} provides a concise comparison across major design aspects.

\begin{table*}[t]
\centering
\caption{Comparison Between AutoFlows and AutoFlows++}
\label{tab:comparison}
\begin{tabular}{lcc}
\hline
\textbf{Feature} & \textbf{AutoFlows} & \textbf{AutoFlows++ (This Work)} \\
\hline

Mining Strategy &
Global causality inference &
Hierarchical local and global mining \\

Trace Processing &
Whole-trace analysis &
Interface-sliced + Whole-trace analysis\\

Causality Filtering &
Confidence-driven edge pruning &
Confidence-driven + local validation \\


Flow Selection &
Heuristic refinement process &
Energy-based optimization framework \\

Temporal Modeling &
None &
Explicit positional constraints \\

Hierarchical Mining &
None &
Local + Global mining \\


Model Construction &
Iterative refinement process &
Single-pass incremental construction \\

Scalability &
Poor scalability &
High scalability with linear traversal \\


\hline
\end{tabular}
\end{table*}
\section{Preliminaries And Problem Formulation}
\label{sec:preliminaries}

\begin{figure}[tb]
\begin{center}
\centering
\begin{tabular}{p{1in}p{1.4in}}
\begin{minipage}{1in}
\centering
\includegraphics[width=.8in]{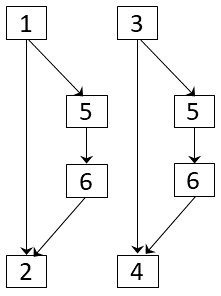}
\end{minipage}
&
\begin{minipage}{1in}
\centering
\begin{small}
\begin{verbatim}
1 (cpu0:cache:rd:req)
2 (cache:cpu0:rd:resp)
3 (cpu1:cache:rd:req)
4 (cache:cpu1:rd:resp)
5 (cache:mem:rd:req)
6 (mem:cache:rd:resp)
\end{verbatim}
\end{small}
\end{minipage}

\\ 
\centering
(a)
& 
\;\;\;\;\;\;\;\;\;(b)
\\
\multicolumn{2}{c}{\begin{small}\texttt{Initial messages = \{1,3\}}\end{small}}
\\
\multicolumn{2}{c}{\begin{small}\texttt{Terminal messages = \{2,4\}}\end{small}}
\\
\multicolumn{2}{c}{\centering (c)} 
\end{tabular}
\caption{(a) Graph representation of CPU downstream flows from Fig.~\ref{fig:soc-ex}(b), where nodes are labeled with messages defined in (b). (c) shows the flows' initial and terminal messages.}
\label{fig:flow-ex}
\end{center}
\vspace*{-20pt}
\end{figure}

This section provides the necessary background for the proposed method, following the formulation introduced in~\cite{OurTCAD}. 
Message flows serve as a formal representation of communication protocols spanning multiple interacting components in a SoC.

\subsection{Communication Trace and Causal Relation}
A message flow captures ordered communication dependencies among system components. 
Fig.~\ref{fig:soc-ex}(b) illustrates an example message flow corresponding to a downstream memory read operation that involves CPU, cache, and memory components while excluding cache-coherence behavior. 
In this work, message flows are modeled as directed acyclic graphs (DAGs), as illustrated in Fig.~\ref{fig:flow-ex}.
Each message is represented as a quadruple $(src:dest:cmd:type)$ where the fields respectively denote the source component, destination component, operation performed at the destination, and whether the message corresponds to a request or response. 
For example, the message $(cpu0:cache:rd:req)$ represents a read request issued by ${\tt cpu0}$ to the ${\tt cache}$.
Message flows begin with an initial message that starts a flow instance and terminate with completion messages indicating the end of a transaction. 
Different execution scenarios correspond to alternative paths within the flow graph. 
As shown in Fig.~\ref{fig:flow-ex}(a), a CPU memory read may follow either a cache-hit sequence $(1,2)$ or a cache-miss sequence $(1,5,6,2)$.

Execution of message flows in real systems is inherently concurrent. 
Multiple flow instances may execute simultaneously, producing traces formed through interleaving. 
We represent an execution trace as $\rho=(m_0, m_1, \cdots, m_n)$ where each $m_i$ denotes a message occurrence. 
If $i < j$, the ordering relation is written as $m_i <_p m_j$.
For instance, if ${\tt CPU0}$ executes its memory-read flow three times while ${\tt CPU1}$ executes two instances, one possible trace is 
\begin{equation}
\label{eq:ex-trace}
(3,1,1,5,4,6,2,5,6,2,1,2,3,4).
\end{equation} 
Although ordering exists in the trace, it does not directly reveal which messages belong to the same flow instance, creating the fundamental challenge addressed by message flow mining.

Message flows represent sequences of causality relations among messages. 
These relations originate from a broader concept known as structural causality, based on the observation that each message generated by a component is typically a response to an earlier input message.

\begin{definition}
\label{def:causal}
Message $m_j$ is \emph{causal} to message $m_i$, denoted as $\mathit{causal}(m_i, m_j)$, if $m_i{\tt .dest} = m_j{\tt .src}.$
\end{definition}
Causal messages mined from traces are also referred to as binary patterns; these two terms are used interchangeably throughout this paper.

Structural causality captures connectivity constraints imposed by component interfaces but does not necessarily represent true protocol dependencies. 
Functional or true causality forms a subset of structural causality and reflects genuine transaction relationships among messages. 
The goal of message flow mining is therefore to distinguish true causal relations from structurally possible ones using observed execution traces.

\subsection{Message Flows and Candidate Paths}

Binary relations can be combined to form longer sequences that represent candidate communication flows.
A candidate path $P$ is defined as an ordered sequence of binary patterns: $P=\{m_1,m_2,\cdots,m_k\}$ where the destination message of $m_i$ matches the source message of $m_{i+1}$ for $1\leq i < k$.
Candidate paths represent potential communication flows that may occur within the system.

A message flow instance corresponds to the occurrence of a candidate path in the execution trace.
Given a candidate path $P$, a flow instance is a sequence of messages such that the messages correspond to the binary patterns defined in $P$, and the messages preserve the ordering constraints of the execution trace.
Flow instances represent concrete realizations of communication flows observed during system execution.

\textbf{Example:} We revisit the motivating example from AutoFlows \cite{AutoFlows} to illustrate the difficulty of this task.
Consider the trace from \ref{eq:ex-trace} and the candidate flows derived from structural causality: $\{1,2\}, \{1,5,6,2\}, \{3,4\}, \{3,5,6,4\}$.
From prior knowledge of the trace in (\ref{eq:ex-trace}), we know that it contains one occurrence of the message flow ${1,2}$, two occurrences of ${1,5,6,2}$, and one occurrence of ${3,4}$, with these flows interleaved within the trace.

Messages create flow instances as they appear in the trace. 
Under heavy interleaving, multiple active instances may compete for the same message occurrence. 
In this example, the message $5$ can extend either the flow $\{1,5,6,2\}$ or $\{3,5,6,2\}$. 
A purely sequential matching policy assigns the message to one instance, leaving the other unfinished and reducing explainability despite the existence of a valid decomposition.
For example, if the first occurrence of message $5$ is incorrectly matched to the first occurrence of message $3$, one instance of the flow ${1,5,6,2}$ remains incomplete, preventing its correct extraction during the mining process.
This example highlights a key limitation of global inference based solely on structural causality: locally valid edges may combine into globally inconsistent flows.

AutoFlows++ addresses this challenge through three complementary mechanisms. 
First, traces are decomposed into \emph{interface slices}, exposing locally consistent communication behavior between component pairs. 
Second, candidate flows are evaluated using a \emph{global path energy model}, where edge confidence and positional indexing jointly determine flow plausibility.
Finally, \emph{positional indexing} augments each message with its trace position, resolving assignment ambiguity by enforcing temporal ordering constraints. 

To evaluate the quality of inferred message flows, we adopt the \emph{acceptance ratio} (Equation~\ref{eq:acceptanceRatio}) from \cite{OurTCAD}, which measures the proportion of messages in the trace that can be explained by the inferred flows.
Formally, given a trace $\rho$ and a set of inferred flows $\mathcal{M}$, the acceptance ratio is defined as the fraction of messages in $\rho$ that can be assigned to valid flow instances in $\mathcal{M}$.
A higher acceptance ratio indicates that the inferred flows better explain the observed system behavior, while a lower acceptance ratio suggests incomplete or incorrect flow reconstruction.

\begin{equation}
\label{eq:acceptanceRatio}
AR = \frac{\sum_{\rho \in T} |\mathrm{Accepted}(\rho)|}{\sum_{\rho \in T} |\rho|}
\end{equation}

Formally, given a set of execution traces, the objective of AutoFlows++ is to infer a compact set of message flows that explains observed behavior with high acceptance ratio while minimizing invalid path combinations introduced by interleaving.

\section{Method}
\label{sec:method}

\begin{figure}[ht]
    \centering
        \includegraphics[width=1\columnwidth]{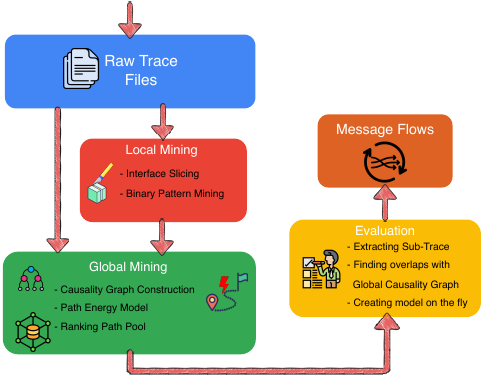}
    \caption{Overview of the AutoFlows++ Framework}
    \label{fig:overallFramework}
\end{figure}

Fig.~\ref{fig:overallFramework} illustrates the overall workflow of the proposed \textbf{AutoFlows++} framework. 
The method takes as input a set of system communication traces and begins with a local mining stage, which extracts binary patterns (BPs) from interface-sliced traces to capture localized interaction dependencies among system components. It then proceeds to a global mining stage, where a global causality graph is constructed and a path energy model is used to enumerate and rank candidate paths. The BPs identified during local mining are leveraged to guide and constrain this global analysis. In the final phase, the system model is incrementally constructed on-the-fly during evaluation. By eliminating the iterative refinement stage and adopting a single-pass traversal of traces, the proposed approach achieves substantially lower computational overhead compared to the original AutoFlows framework \cite{AutoFlows}.
Algorithm~\ref{algo:overall-algorithm} provides the procedural implementation corresponding to the workflow illustrated in Fig.~\ref{fig:overallFramework}.
The following subsections describe each step in detail.

\begin{algorithm}[tb]
  \SetKwInOut{Input}{input}
  \SetKwInOut{Output}{output}
\caption{\textbf{Proposed Method}}
\label{algo:overall-algorithm}
  \Input{A set of traces $\mathcal{T}$}
  \Output{Message flow model mined from $\mathcal{T}$}
  \Output{Acceptance Ratio $\mathit{AR}$}
  $\mathcal{BP_V}, \mathcal{BP_I}, \mathcal{MBP} = {\tt LocalMining}(\mathcal{T})$\;
  $\mathcal{RPP} = {\tt GlobalMining}(\mathcal{T}, \mathcal{BP_V}, \mathcal{BP_I}, \mathcal{MBP})$\;
  $\mathit{AR}, \mathit{Model}  = {\tt PosAwareEval}(\mathcal{T}, \mathcal{RPP}, \mathcal{MBP})$\; 
  \Return $\mathit{Model}$, $\mathit{AR}$\;
\end{algorithm}
\setlength{\textfloatsep}{0pt}

\subsection{Local Mining}
\label{subsec:localMining}

The first stage of AutoFlows++ performs local mining, which extracts binary message relations from execution traces projected onto individual interfaces between pairs of system components. The goal of this stage is to identify locally consistent causal relationships while filtering out spurious relations introduced by concurrency and message interleaving. The local mining process consists of two main steps: interface slicing, followed by binary pattern validation, which includes binary pattern confidence computation and minimal pattern selection.

\subsubsection{Interface Slicing}
Components in a system design communicate with one another through well-defined interfaces to realize system-level transactions. 
Instead of mining binary relations from the entire execution trace, AutoFlows++ first partitions the trace according to the interfaces between components. 
Mining is then performed independently on the resulting interface-specific traces.

Given an execution trace $\rho=\{m_0,m_1,\cdots,m_n\}$, each message $m_i$ belongs to an interface determined by its source and destination attributes. 
The trace can therefore be decomposed into a set of interface-specific traces:
\[
\rho_{(s,d)}=\{m_i \in \rho | m_i.src=s \lor m_i.dest=d\} 
\]
where each sliced trace captures the communication behavior between two components $s$ and $d$.

The procedure for interface slicing is described in Lines 2–4 of Algorithm~\ref{algo:localMining}. 
The algorithm scans the execution trace and assigns each message to its corresponding interface-specific trace. 
This decomposition isolates local communication patterns and enables binary relations to be mined independently for each interface.

\subsubsection{Binary Pattern Validation}
\label{subsubsec:bpValidation}

After interface slicing, AutoFlows++ performs binary pattern validation within each interface-sliced trace. 
The objective of this step is to identify reliable causal relations between messages while filtering out spurious relations introduced by concurrent execution. 
The process begins by extracting all candidate binary patterns, which correspond to causal message pairs defined in Definition~\ref{def:causal}. 
For each candidate binary pattern denoted as $A \rightarrow B$, two statistical confidence measures are computed. 
The forward confidence quantifies how frequently message $A$ is followed by message $B$.
\begin{definition}
\label{def:forwardConfidence}
Forward-confidence of edge $A \to B$:
\begin{equation}
\label{eq:forwardConfidence}
ForwardConfidence(A, B)=\frac{freq(A \rightarrow B)}{freq(A)}
\end{equation} 
\end{definition}

The backward confidence measures how frequently message $B$ is preceded by message $A$:
\begin{definition}
\label{def:backwardConfidence}
Backward-confidence of edge $A \to B$:
\begin{equation}
\label{eq:backwardConfidence}
BackwardConfidence(A, B)=\frac{freq(A \rightarrow B)}{freq(B)}
\end{equation} 
\end{definition}
In the above definitions, $freq(A \rightarrow B)$ denotes the number of occurrences in which message type $A$ is followed by message type $B$ in the interface slice.
These confidence values quantify the strength of the causal relationship between the two messages. 
The procedure for computing the confidence values is described in lines 5–8 of Algorithm~\ref{algo:localMining}.

Typically, a number of messages are exchanged over an interface during system execution.
Among these messages, causal relations may be valid or invalid. 
AutoFlows++ aims to identify a subset of causal relations that can accurately explain the observed communication behavior on an interface.
Specifically, the algorithm identifies a minimal set of binary patterns that cover all messages in an interface-specific traces with the maximal confidence scores. 
These identified binary patterns are referred to as \emph{valid} patterns.
Formally, given a set of candidate patterns $P$, the selection procedure determines a subset $P^*$ such that:

\begin{itemize}
    \item Every message in the interface-specific trace participates in at least one selected binary pattern.
    \item The selected patterns maximize forward and backward confidence values.
    \item The number of selected patterns is minimized.
\end{itemize}
The last requirement is based on the Occam’s razor \cite{OccamsRazor} theorem, simpler models, smaller $P^\ast$ in this case, typically provide better generalization and improved interpretability.

Binary patterns that are not selected are classified as invalid. 
As a result, this step produces three outputs: a set of valid binary patterns, a set of invalid binary patterns, and a mapping of all matched binary pattern instances, as described in Algorithm~\ref{algo:localMining}, lines 9–11.

Matching binary patterns refers to the process of associating each source message instance of a valid binary pattern with its corresponding destination message instance. 
For each such match, the original position of the destination message in the unsliced trace is recorded for subsequent use in the evaluation procedure.

\textbf{Example:} To illustrate the advantages of AutoFlows++ over AutoFlows \cite{AutoFlows}, we consider a simple example based on the message definitions in Fig.~\ref{fig:flow-ex}.
Consider the following execution trace:
\begin{equation}
\label{eq:advTrace}
(1,3,5,2,6,4,1,5,6,2,3,4).
\end{equation}


Based on the ground-truth specification, the trace contains one instance of each message flow $\{1,2\}$, $\{3,4\}$, $\{1,5,6,2\}$, and $\{3,5,6,4\}$, interleaved within the execution. 
The same example trace is reused in the following subsection.

The trace can be decomposed into three communication interfaces: {\tt \{cpu0-cache\}}, {\tt \{cpu1-cache\}}, and {\tt \{cache-mem\}}. 
Based on this decomposition, the corresponding interface-sliced traces are $\{1,2,1,2\}$, $\{3,4,3,4\}$, and $\{5,6,5,6\}$, respectively. 
From these slices, the binary patterns $\{1,2\}$, $\{3,4\}$, and $\{5,6\}$ are identified as valid based on their high forward and backward confidence values. 
In this simplified example, no invalid binary patterns are observed.






\begin{algorithm}[tb]
  \SetKwInOut{Input}{input}
  \SetKwInOut{Output}{output}
\caption{\textbf{Local Mining}}
\label{algo:localMining}
\IncMargin{1.5em}

\Input{A set of traces $\mathcal{T}$}
\Output{Valid and invalid binary patterns $\mathcal{BP}_V$, $\mathcal{BP}_I$}
\Output{Matched binary patterns $\mathcal{MBP}$}

$\mathcal{I} \leftarrow \emptyset$,
$\mathcal{CR} \leftarrow \emptyset$, 
$\mathcal{BP}_V \leftarrow \emptyset$, 
$\mathcal{BP}_I \leftarrow \emptyset$, 
$\mathcal{MBP} \leftarrow \emptyset$ \;
\ForEach{trace $\rho \in \mathcal{T}$}
{
    \ForEach{message instance $m_i \in \rho$}
    {
        Add message $m_i$ to the corresponding interface slice $\mathcal{I}_j$ based on its source and destination\;
    }
}

\ForEach{$\mathcal{I}_i \in \mathcal{I}$}
{
    $CR_i \leftarrow$ find all causal relations in $\mathcal{I}_i$\;
    \ForEach{$bp \in CR_i$}
    {
        $(FC_i, BC_i) \leftarrow$ measure the forward and backward confidence\;
    }
    $\mathcal{BP}_V \leftarrow \mathcal{BP}_V \cup$ minimal set of binary patterns covering all messages in $S_i$ with the highest $FC+BC$\;
    $\mathcal{BP}_I \gets \mathcal{BP}_I \cup (CR_i - \mathcal{BP}_V$)\;
    $\mathcal{MBP}_i \gets$ match each binary pattern instance with the corresponding pattern in $\mathcal{BP}_V$ and record the destination position for each source\;
}

\Return $\mathcal{BP}_V, \mathcal{BP}_I, \mathcal{MBP}$\;
\end{algorithm}

\subsection{Global Mining}
\label{subsec:globalMining}

Following the local mining stage, which identifies reliable binary communication patterns within individual interface slices, AutoFlows++ conducts a global mining stage to infer system-level message flows that span multiple components. 
The objective of this phase is to aggregate the locally mined binary relations into candidate message flows and to prioritize these flows according to an energy  model. 

Global mining comprises two steps: (1) construction of a causality graph, and (2) ranking of paths using a path energy model.

\subsubsection{Causality Graph Construction}
To construct causality graphs, the first step is to extract unique messages from the input traces $T$, where a unique message is defined as one that differs in at least one attribute from previously collected messages. 
We assume that users specify the initial and terminal messages of interest. 
Let $M$ denote the set of all unique messages extracted from $T$, with the sets of initial and terminal messages denoted by $i$ and $t$, respectively, where $i \subseteq M$ and $t \subseteq M$.

A causality graph is defined as a directed acyclic graph (DAG) with multiple roots and terminal nodes. 
Each node in the graph corresponds to a unique message from the set $M$, where root nodes represent initial messages from $i$ and terminal nodes correspond to messages in $t$. 
Edges represent structural causality relations between messages.

For each initial message $m_i \in i$, a causality graph is constructed by first creating a root node for $m_i$, followed by adding nodes for all messages $m$ such that $causal(m_i, m)$ holds. 
This process is applied recursively until terminal messages are reached. 
Edges that would introduce cycles are omitted to preserve the acyclic property of the graph.
Each path from a root node to a terminal node represents a potential message flow.
Given that the number of paths in a causality graph can be large, it is necessary to identify and filter out paths that correspond to likely invalid message flows. 

After constructing the causality graph, edges corresponding to binary patterns classified as invalid during the local mining step are removed. 
For edges belonging to the set of valid binary patterns, the confidence values obtained during the local mining step (from interface-sliced traces) are reused. 
For the remaining edges, the confidence values are computed using the global trace (i.e., the unsliced trace), as defined in Definitions~\ref{def:forwardConfidence} and~\ref{def:backwardConfidence}.

\begin{figure}
    \centering
    \begin{tabular}{ccc}
         \begin{minipage}{2in}
             \includegraphics[width=2in]{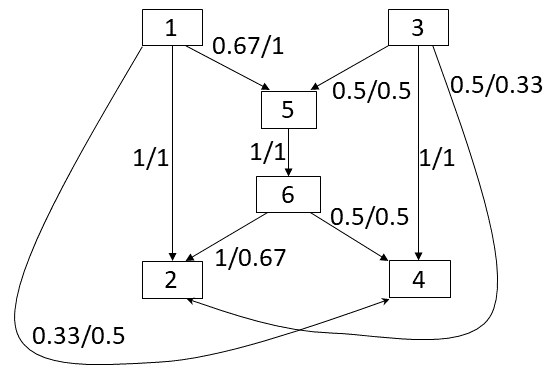}
        \end{minipage}
    \end{tabular}
    \caption{Causality graph from trace~(\ref{eq:ex-trace}): edges marked with forward/backward confidences.}
    \label{fig:cg-supp-conf-ex}
\end{figure}

Given a set of traces $T$, the final forward and backward confidence values are obtained by averaging the corresponding confidence values computed for each trace $\rho \in T$. 
Fig.~\ref{fig:cg-supp-conf-ex} shows the causality graph for trace~(\ref{eq:ex-trace}) annotated with forward and backward confidence values. 
This information is used to guide the global mining.

\subsubsection{Path Energy Model}
In path energy model step we first extract all available paths from initial messages to the terminal messages from the constructed causality graph and call them candidate paths.
To rank candidate paths, AutoFlows++ defines a Path Energy Model that characterizes each candidate flow through a unified cost that captures structural validity, relation strength, and temporal alignment with the execution trace.
Given a candidate path $P$, the energy of the path is computed using three factors:
\begin{itemize}
    \item [1.] Confidence of binary relations.
    \item [2.] Validity of binary patterns.
    \item [3.] Distance between messages in the trace.
\end{itemize}

Formally, the energy of a candidate path $P$ is defined as:
\begin{equation}
\label{eq:pathEnergyModel}
\begin{aligned}
E(P) = 
& \frac{1}{|P|} \sum_{e_i \in P} -\log \mathrm{Conf}(e_i) \\
& -\log \left( \frac{VBP_{e_i}}{|P|} \right) \\
& + \frac{1}{|P|} \sum_{e_i \in P} d(e_i)
\end{aligned}
\end{equation}

where $Conf(e_i)$ denotes the mean of the forward and backward confidence values of edge $e_i$, $VBP_{e_i}$ represents the number of valid binary patterns associated with $e_i$, and $\frac{1}{|P|}\sum_{e_i \in P} d(e_i)$ represents the mean distance of the binary patterns present along the path.


The first term reflects the statistical strength of the binary relations composing the path. 
Paths consisting of high-confidence relations therefore receive lower energy values. 
The second term favors paths that incorporate a larger number of validated binary patterns discovered during the local mining stage, thereby increasing the robustness of the inferred flow. 
The third term captures the temporal consistency of the path by measuring the positional distance between messages in the execution trace, rewarding paths whose constituent messages occur with consistent spacing.

Paths with lower energy values are therefore considered more likely to represent valid communication flows.
The computation of path energy and the ranking of candidate paths are described in Algorithm~\ref{algo:globalMining} (Global Mining).
The path energy can be interpreted as a composite cost that penalizes weak causal relations, unsupported binary patterns, and inconsistent temporal spacing.

\textbf{Example:} In the global mining stage, the causality graph is constructed based on the trace in (\ref{eq:advTrace}). 
The resulting graph has the same structural topology as the one shown in Fig.~\ref{fig:cg-supp-conf-ex}, with differences only in the forward and backward confidence values associated with the edges.

Based on this causality graph, all candidate paths are enumerated, and their corresponding path energies are computed according to the path energy model. 
The ranked path pool and their associated path energies can be found in (\ref{eq:pathEnergy}).
\begin{equation}
\label{eq:pathEnergy}
\begin{array}{l c l}
\text{Path} & \rightarrow & \text{Path Energy} \\
\{3,4\}     & \rightarrow & 2.5    \\
\{1,5,6,4\} & \rightarrow & 2.76   \\
\{1,2\}     & \rightarrow & 3      \\
\{1,5,6,2\} & \rightarrow & 3.09   \\
\{3,5,6,4\} & \rightarrow & 3.33   \\
\{3,5,6,2\} & \rightarrow & 3.66   \\
\{1,4\}     & \rightarrow & \infty \\
\{3,2\}     & \rightarrow & \infty \\
\end{array}
\end{equation}





\begin{algorithm}[tb]
  \SetKwInOut{Input}{input}
  \SetKwInOut{Output}{output}
\caption{\textbf{Global Mining}}
\label{algo:globalMining}
\IncMargin{1.5em}

\Input{A set of traces $\mathcal{T}$}
\Input{Valid and Invalid Binary Patterns $\mathcal{BP}_V$, $\mathcal{BP}_I$}
\Input{Matched Binary Patterns $\mathcal{MBP}$}
\Output{Ranked Path Pool $\mathcal{RPP}$}

$\mathcal{CG} \leftarrow {\tt ConstructCausalityGraph}(\mathcal{T})$\;
$\mathcal{CG} \leftarrow {\tt RemoveEdges}(\mathcal{CG}, \mathcal{BP}_I)$\;
$\mathcal{PP} \leftarrow$ find all available paths from initial nodes to terminal nodes in $\mathcal{CG}$\;
$\mathcal{RPP} \leftarrow \emptyset$\;

\ForEach{$p, FC, BC \in \mathcal{PP}$}
{
    $\mathcal{RPP} \leftarrow \mathcal{RPP} \cup \{{\tt PathEnergyModel}(p, FC, BC, \mathcal{BP}_V, \mathcal{BP}_I, \mathcal{MBP})\}$\;
}

$\mathcal{RPP} \leftarrow {\tt Sort}(\mathcal{RPP}, \text{increasing})$\;

\Return $\mathcal{RPP}$\;

\DecMargin{1.5em}
\end{algorithm}



\subsection{Model Evaluation}
After the completion of local and global mining, the framework proceeds to the model evaluation stage, during which the final message-flow model is constructed incrementally. 
In this phase, the matched BPs extracted in the local mining stage and the ranked candidate paths obtained from global mining jointly guide the acceptance and integration of message sequences into the inferred model.

The evaluation algorithm scans each trace sequentially to identify candidate initial messages, which serve as potential starting points for valid message flows. 
Messages that are not preceded by an initial event are immediately marked as invalid, since all mined flows must originate from a valid initial condition. 
For every detected initial message, the algorithm locates the corresponding terminal message and extracts the bounded sub-trace spanning from the initial to the terminal event. 
This localized processing strategy, referred to as position-aware evaluation, preserves the temporal ordering of events and constrains the analysis to execution segments that are structurally feasible.



A sub-causality graph is constructed from each extracted sub-trace to capture feasible causal relations within the localized segment, following the same construction procedure as in section \ref{subsec:globalMining} ({\tt ConstructCausalityGraph}$(SubT)$, Algorithm~\ref{algo:evaluation-algorithm}, line 8). 
All candidate paths are then enumerated and compared against the ranked path pool derived from the global causality graph. 
The algorithm selects the longest candidate path that overlaps with a globally mined path while minimizing the associated path energy.
Upon identifying a candidate match, the algorithm evaluates whether removing the corresponding messages from the graph results in orphan nodes.
This step reduces the likelihood of selecting paths that include messages not belonging to the current flow instance, but instead associated with other flows occurring later in the trace.
If orphan nodes are introduced, alternative candidate paths are considered. 
When no candidate path can eliminate orphan nodes entirely, the algorithm selects the path that minimizes the number of remaining orphan nodes ({\tt LowestEnergyWithMinOrphanNodes}, Algorithm~\ref{algo:evaluation-algorithm} line 10).
The messages along the selected path are then marked as accepted and removed from further consideration in the trace. 
Additionally, the corresponding path is incorporated into the inferred model if it has not been previously included, enabling incremental model construction.

This procedure is repeated for subsequent initial messages until all events in the trace have been examined. 
By combining locally validated binary patterns with globally ranked path structures, the evaluation stage facilitates robust and consistent flow extraction while maintaining computational efficiency through localized sub-trace analysis and single-pass incremental model construction.

Incorporating positional awareness provides several important benefits. 
First, it prevents associations of messages during the model evaluation over the original traces that  would be identified as invalid during the local mining stage. 
This makes the results of model evaluation on original traces consistent with results obtained during the local mining stage.
Second, it reduces ambiguity when multiple active flow instances compete for the same binary pattern occurrence, a common scenario under heavy interleaving of concurrent transactions. 
Third, positional indexing improves the correctness and explainability of the inferred flows, as the resulting assignments better reflect the true temporal relationships among messages. 
Importantly, because the proposed position-aware evaluation strategy processes each trace in a single pass and eliminates the need for iterative refinement cycles, the overall runtime is substantially lower than that of the original AutoFlows framework \cite{AutoFlows}. 
This design choice enables efficient model construction while maintaining robust flow inference grounded in both local and global structural evidence.

\begin{figure}[ht]
    \centering
        \includegraphics[width=1\columnwidth]{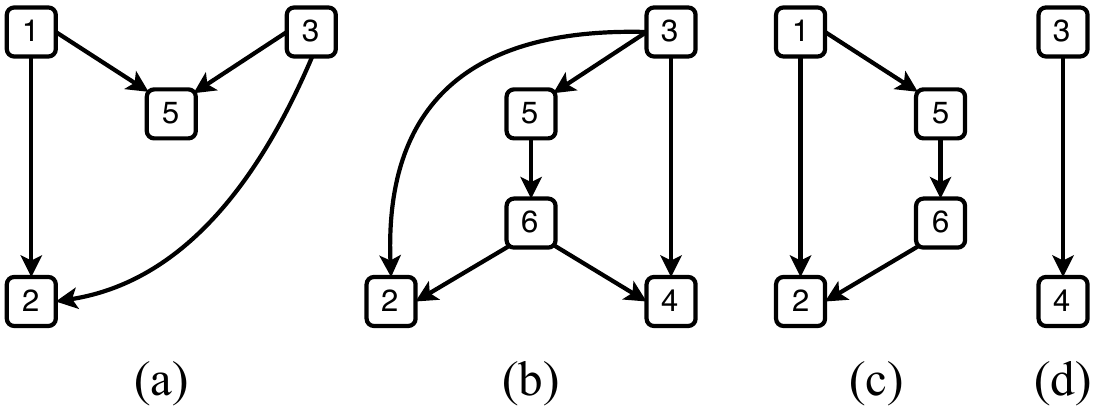}
    \caption{Sub-causality graphs for the example trace~\ref{eq:advTrace}.}
    \label{fig:subCausalityGraphs}
\end{figure}

\textbf{Example:} During the evaluation phase, AutoFlows++ processes the trace using position-aware sub-trace extraction. 
The first sub-trace is $SubT_1 = \{1,3,5,2\}$ (corresponding sub-causality graph in Fig.~\ref{fig:subCausalityGraphs}(a)), from which the path $\{1,2\}$ is selected. 
The next sub-trace, $SubT_2 = \{3,5,2,6,4\}$ (corresponding sub-causality graph in Fig.~\ref{fig:subCausalityGraphs}(b)), yields the path $\{3,5,6,4\}$. 
Subsequently, $SubT_3 = \{1,5,6,2\}$ (corresponding sub-causality graph in Fig.~\ref{fig:subCausalityGraphs}(c)) produces the path $\{1,5,6,2\}$, and finally, $SubT_4 = \{3,4\}$ (corresponding sub-causality graph in Fig.~\ref{fig:subCausalityGraphs}(d)) yields the path $\{3,4\}$. 
These selections are guided by the ranked path pool and refined using the orphan-node criterion, enabling consistent and complete reconstruction of flow instances.

In contrast, AutoFlows \cite{AutoFlows}, which relies on heuristic global matching without positional awareness, fails to consistently resolve ambiguity under interleaving. 
As a result, it can achieve at most a $66\%$ acceptance ratio on this trace due to incomplete flow reconstruction. 
In comparison, AutoFlows++ achieves a $100\%$ acceptance ratio by leveraging positional indexing to partition the trace into semantically consistent sub-traces, thereby enabling accurate assignment of messages to their corresponding flow instances.

\begin{algorithm}[tb]
  \SetKwInOut{Input}{input}
  \SetKwInOut{Output}{output}
\caption{\textbf{ModelEvaluation}}
\label{algo:evaluation-algorithm}
\IncMargin{1.5em}
  \Input{A set of trace $\mathcal{T}$ with the mapped corresponding response position}
  \Input{Ranked Path Pool $\mathcal{RPP}$}
  \Output{Acceptance Ratio $AR$}
  \Output{Model $\mathcal{M}$}

$SumAR = 0$, $ActivePaths \leftarrow \emptyset$\;
\ForEach{$\rho \in \mathcal{T}$}
{
    \For{$0 \leq i \leq {|\rho|}$}
    {
        $m_i \gets \rho[i]$\;
        $m_t \gets \rho[t]$ such that $t$ is the position of the matched message from local mining\;
        \If{$m_i$ is an initial message, $m_t$ is a terminal message }
        {
            let $SubT$ be the subtrace $\rho[i \ldots j]$\;
            $SubCG \leftarrow {\tt ConstructCausalityGraph}(SubT)$\;
            $pp \leftarrow \text{find all available paths in } {SubCG}$\;
            $CandPath \leftarrow {\tt LowestEnergyWithMinOrphanNodes}(pp \cap \mathcal{RPP})$\;
            \If{$CandPath \text{ exists}$}
            {
                Add all messages in $CandPath$ \mbox{to} $Accepted$\;
                $\mathcal{M} \leftarrow \mathcal{M} \cup CandPath$\;
                Remove all messages in $CandPath$ from $\mathcal{T}$\;
            }
        }
        \Else
        {
            Add $m_i$ \mbox{to} $UnAccepted$\;
        }
    }
    $SumAR = SumAR + |Accepted|/|\rho|$\;
}
\Return $AR = SumAR/|T|$, $\mathcal{M}$\;
  
\DecMargin{1.5em}
\end{algorithm}

\section{Experimental Results}
\label{sec:experimentalResults}
This section evaluates the effectiveness, accuracy, and scalability of the proposed AutoFlows++ framework. 
We conduct experiments using both synthetic communication traces with known ground-truth specifications and architectural traces generated from realistic SoC models simulated in GEM5 \cite{gem5}.
The evaluation focuses on the following research questions:
\begin{itemize}
    \item [] \textbf{RQ1. Accuracy:} How accurately can AutoFlows++ reconstruct message flows under heavy concurrency?
    \item [] \textbf{RQ2. Model Quality:} Does AutoFlows++ produce compact and interpretable communication models?
    \item [] \textbf{RQ3. Scalability:} How does the framework scale with increasing trace length and concurrency level?
    \item [] \textbf{RQ4. Contribution of design components:} What is the impact of hierarchical mining, path energy ranking, and positional indexing?
\end{itemize}

\subsection{Benchmarks and Trace Sources}
We evaluate AutoFlows++ using both synthetic protocol benchmarks and realistic SoC execution traces.

\subsubsection{\textbf{Synthetic Traces}} 
In preliminary experiments, AutoFlows++ is evaluated using synthetic traces generated from ten message flows designed to mimic realistic SoC communication behavior. 
The flows model memory operations and accesses involving CPUs, caches, and peripheral components, capturing communication patterns commonly observed in modern SoC designs.
Three synthetic trace sets are constructed: {\tt small-20}, {\tt large-10}, and {\tt large-20}.
The terms small and large refer to the number of distinct message flows included in trace generation, while the numerical suffix denotes the number of execution instances per flow.
Specifically, {\tt small-20} contains only CPU-initiated flows (4 out of the 10 total flows), each executed 20 times, {\tt large-10} includes all ten flows, each executed 10 times, and {\tt large-20} includes all ten flows, each executed 20 times.

These datasets allow controlled evaluation under varying degrees of concurrency and interleaving intensity. 
The {\tt small-20} benchmark represents limited interaction scenarios dominated by CPU-originated transactions, whereas the {\tt large-*} configurations emulate complex SoC environments where multiple heterogeneous communication flows execute simultaneously.
Because ground-truth message flows are known for synthetic benchmarks, these traces enable precise measurement of reconstruction accuracy, invalid path generation, and explainability improvements introduced by AutoFlows++.

\subsubsection{\textbf{GEM5 Traces}}
\begin{figure}
    \centering
    \includegraphics[width = 0.8\linewidth]{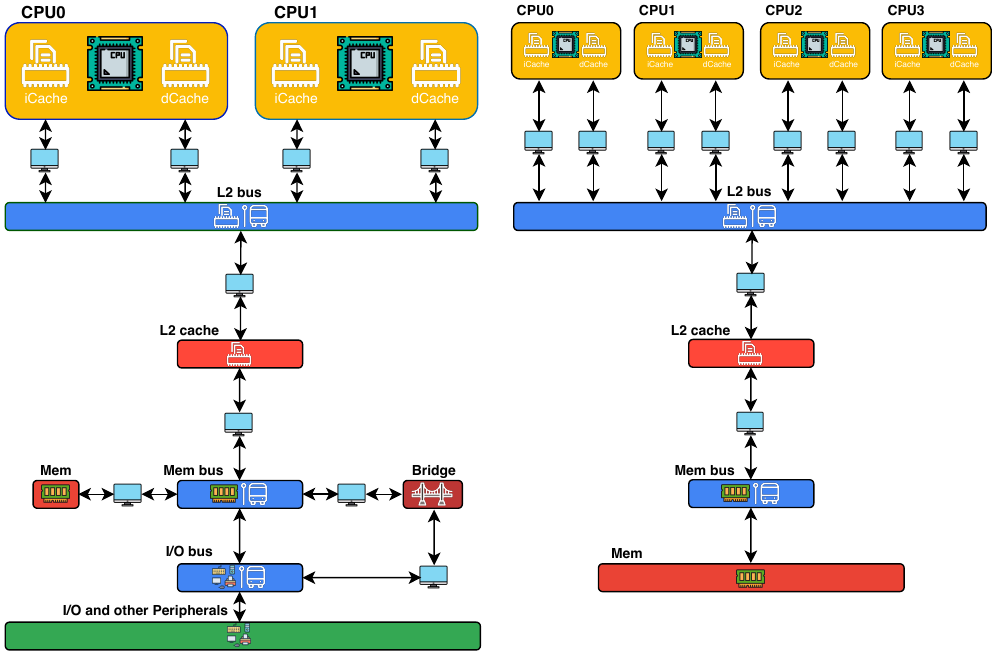}
    \quad \quad\quad \quad \quad (a) \quad \quad \quad \quad \quad \quad \quad \quad \quad (b)
    \caption{(a) GEM5 Full-System (FS) design. (b) GEM5 System Emulation (SE) design. \raisebox{-2.5pt}{\includegraphics[width=0.15in]{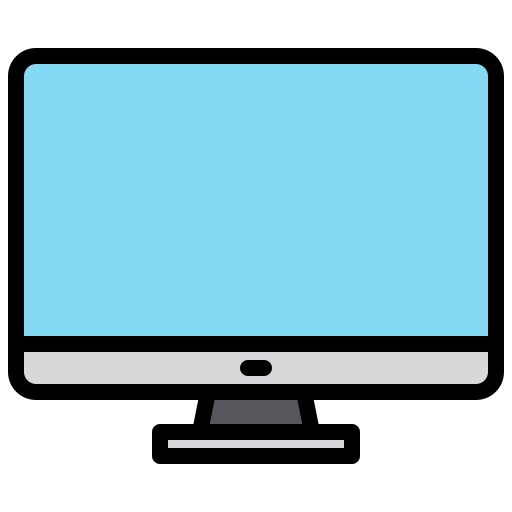}}: Communication Monitor}
    \label{fig:gem5}
\end{figure}

For comprehensive evaluation, we apply AutoFlows++ to execution traces generated using the GEM5 architectural simulator. 
GEM5 supports two execution modes: System-call Emulation (SE) and Full-System (FS), illustrated in Fig.~\ref{fig:gem5}(b) and Fig.~\ref{fig:gem5}(a), respectively.

In SE mode, GEM5 simulates application execution by emulating system calls without modeling a complete operating system. 
This mode requires static thread-to-core mapping and therefore limits simulation of fully dynamic multithreaded environments. 
Despite these constraints, SE mode enables efficient experimentation with application-level workloads.
The SE configuration used in our experiments consists of: four x86 CPU cores, private L1 caches per core, a shared L2 cache, and a 2GB DDR4 main memory.
Two traces collected from the SE setup are {\tt threads}, which is generated from a multithreaded application workload, and {\tt snoop}, which is generated using Peterson’s mutual exclusion algorithm.
These traces capture fine-grained synchronization and cache-coherence interactions arising from concurrent program execution.

In contrast, FS mode simulates a complete operating environment including operating system behavior, interrupts, device interactions, and I/O subsystems. 
Because FS mode models realistic system activity, it provides more accurate execution behavior and exposes a broader range of communication patterns.
The FS configuration includes: two CPU cores, a DDR3 main memory, operating-system services and I/O devices.
The FS trace used in evaluation corresponds to the boot sequence of an Ubuntu 18 Linux operating system, producing complex communication behavior involving memory management, interrupts, and device initialization.

Together, the SE and FS datasets enable evaluation of AutoFlows++ across controlled application-level workloads and realistic system-level execution environments, demonstrating robustness under diverse communication scenarios.

\subsection{Accuracy on Synthetic Benchmarks}
Table~\ref{table:syntheticResults} reports quantitative results obtained on synthetic traces.
Across all synthetic benchmarks, AutoFlows++ achieves substantially higher acceptance ratios than prior approaches.
Specifically, AutoFlows++ reaches acceptance ratios of 99.51\%, 99.31\%, and 98.62\% on Small-20, Large-10, and Large-20 traces respectively, compared with 89.57\%–90.47\% achieved by AutoFlows \cite{AutoFlows} and 57.12\%–69.97\% achieved by AutoModel \cite{OurTCAD}. 
These results demonstrate that hierarchical mining combined with positional reasoning significantly improves the ability to correctly assign messages to flow instances under heavy interleaving.

The improvement is particularly pronounced in highly concurrent settings, where AutoFlows++ outperforms AutoFlows by more than 9 percentage points and AutoModel by over 41 percentage points. 
This gain highlights the effectiveness of interface slicing in eliminating spurious causal relations prior to global inference.

In addition to improved accuracy, Table~\ref{table:syntheticResults} shows that AutoFlows++ produces smaller models in comparison to AutoFlows \cite{AutoFlows}. 
Smaller models are preferable as they provide a more compact and interpretable representation of system behavior, consistent with Occam’s razor \cite{OccamsRazor}, while reducing redundancy and improving generalization.
Moreover, due to the single-pass traversal of traces and the absence of iterative refinement, the proposed method achieves substantially lower runtime, enabling efficient processing of long traces without compromising model accuracy.

\subsection{Accuracy on GEM5 Architectural Traces}
We next evaluate AutoFlows++ on realistic architectural traces obtained from GEM5 simulations \cite{gem5}, including Threads, Snoop, and Full-System workloads.
These traces capture complex system-level interactions involving cache coherence, synchronization primitives, memory hierarchies, and operating-system activity.

As summarized in Table~\ref{table:gem5results}, AutoFlows++ consistently achieves the highest acceptance ratios across all evaluated workloads.
For example, AutoFlows++ reaches 98.91\% acceptance ratio on the Threads trace containing approximately 7.6 million messages, outperforming AutoFlows (97.92\%) \cite{AutoFlows} and AutoModel (96.31\%) \cite{OurTCAD}.
Similarly, on the Snoop trace, the proposed method improves acceptance ratio to 95.74\%, compared with 92.67\% and 88.79\% for AutoFlows and AutoModel respectively.

While the percentage improvements over AutoFlows \cite{AutoFlows} may appear modest in some cases, it is important to consider the absolute scale of the execution traces. 
Even a one-percentage-point increase in acceptance ratio corresponds to tens or hundreds of thousands of additional messages correctly explained by the inferred model. 
For instance, a 1.0\% improvement on the Threads trace corresponds to $\approx 76,000$ additional messages explained.
For large architectural traces, this translates into substantially improved coverage of communication behavior and more reliable reconstruction of protocol interactions.

To further evaluate scalability, we apply AutoFlows++ to a Full-System trace containing approximately $8.4\times 10^9$ messages.
Despite the extreme size and the presence of diverse OS-level activities, the proposed framework achieves an acceptance ratio of 92.15\% while producing a compact model consisting of 198 inferred flows. 
The total mining runtime for this trace is approximately 32 minutes, demonstrating the practical scalability of the proposed hierarchical mining pipeline. 
This represents a substantial improvement over the 9 hours and 46 minutes required by AutoFlows \cite{AutoFlows} and the 4 hours and 11 minutes reported for AutoModel \cite{OurTCAD}.
Compared to AutoFlows, the proposed framework exhibits improved stability in large-scale inference despite the increased complexity of full-system execution. 
In particular, the hierarchical local mining stage effectively reduces spurious causal relations prior to global inference, leading to more consistent and reliable reconstruction of communication behavior across diverse protocol interactions.

These results confirm that the proposed framework scales effectively to realistic SoC workloads while maintaining high reconstruction accuracy.

\begin{figure*}
    \centering
    \begin{subfigure}[b]{0.32\textwidth}
        \includegraphics[width = \textwidth, left]{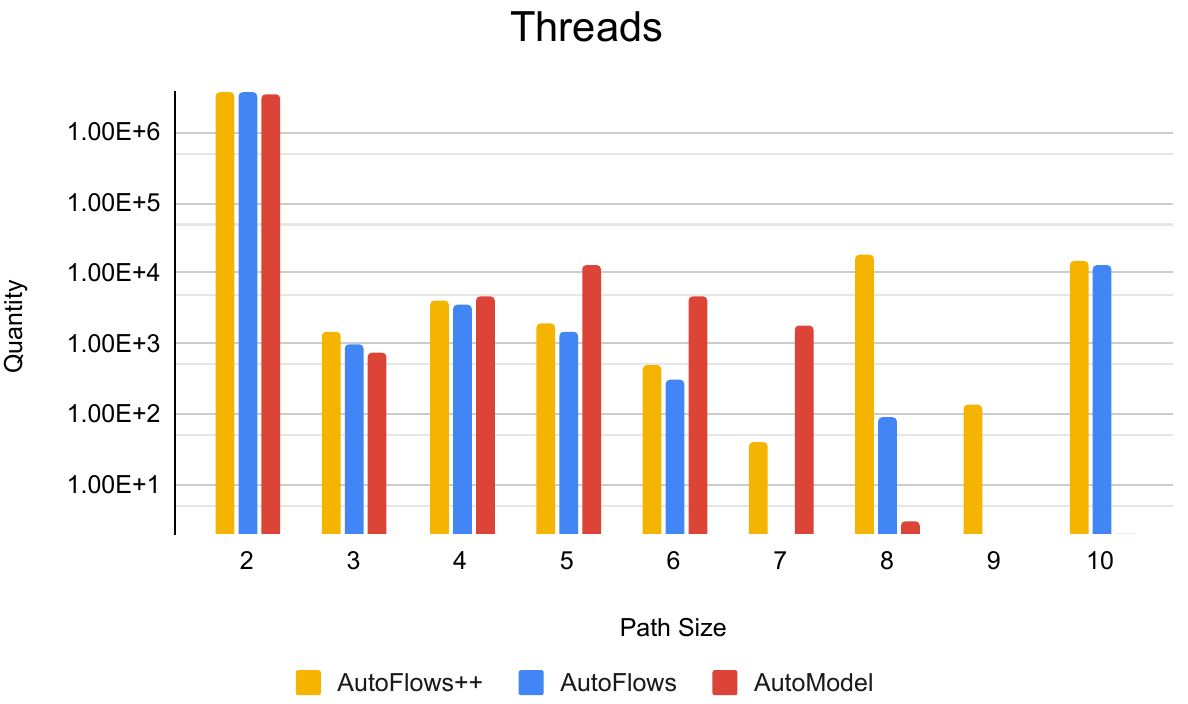}        
        \caption{}
    \end{subfigure}
    \hfill
    \begin{subfigure}[b]{0.32\textwidth}
        \includegraphics[width = \textwidth, right]{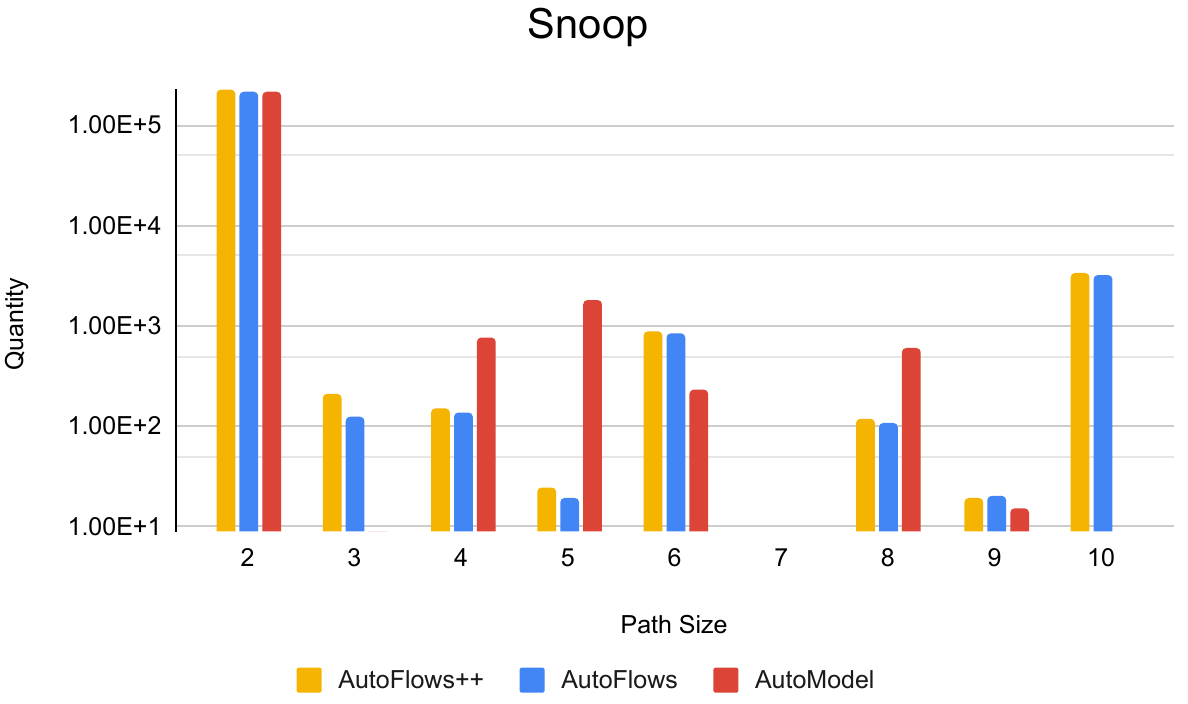}
        \caption{}
    \end{subfigure}
    \hfill
    \begin{subfigure}[b]{0.32\textwidth}
        \includegraphics[width = \textwidth, right]{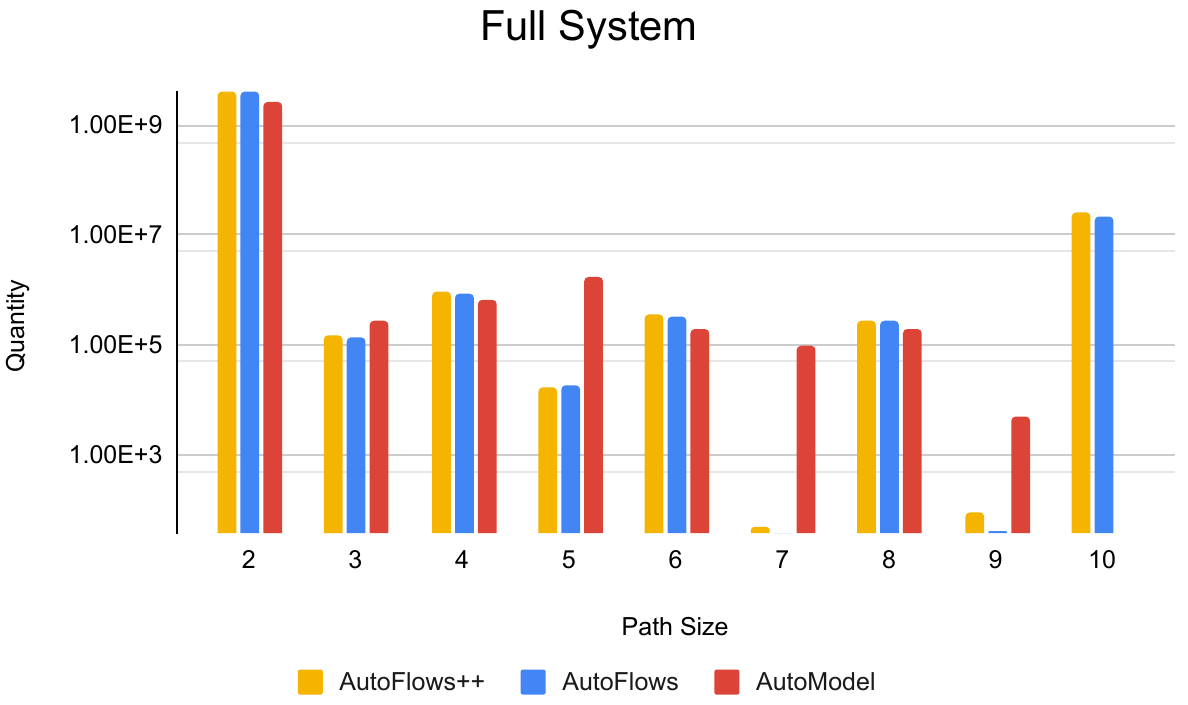}
        \caption{}
    \end{subfigure}
     
    \caption{Counts of instances of the mined message sequences of different lengths using the AutoFlows++, AutoFlows, and AutoModel \iffalse model synthesis\fi method that are found in the GEM5 (a) {\tt threads}, (b) {\tt snoop}, and (c) {\tt full-system} traces.}
    \label{fig:pathQuantity-all}
\end{figure*}
\begin{figure*}
    \centering
    \begin{subfigure}[b]{0.33\textwidth}
        \includegraphics[width = \textwidth, left]{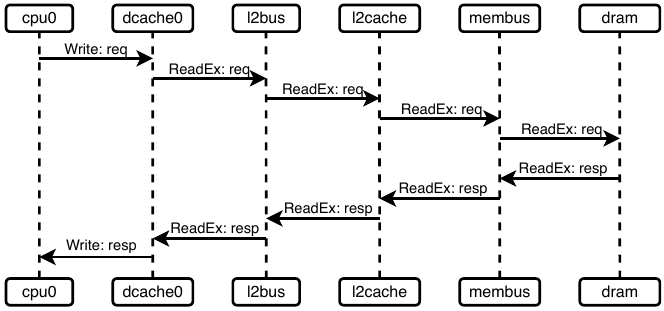}        
        \caption{}
    \end{subfigure}
    \hfill
    \begin{subfigure}[b]{0.305\textwidth}
        \includegraphics[width = \textwidth, right]{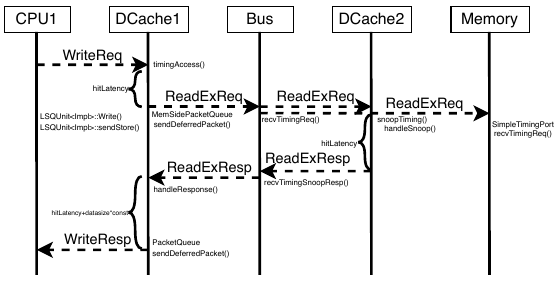}
        \caption{}
    \end{subfigure}
    \hfill
    \begin{subfigure}[b]{0.33\textwidth}
        \includegraphics[width = \textwidth, right]{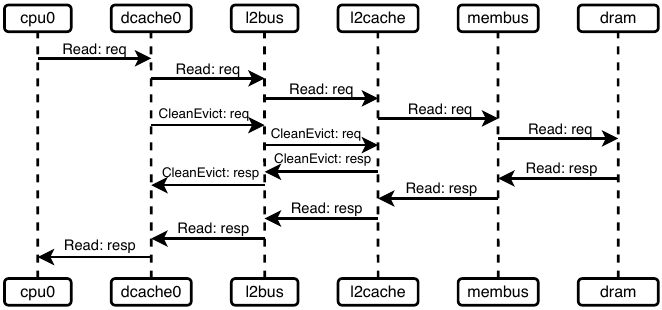}
        \caption{}
    \end{subfigure}
     
    \caption{(a) A flow mined from the GEM5 threads trace. (b) GEM5 documentation's memory write miss sequence. (c) Two flows mined from the GEM5 threads trace depicting a clean eviction scenario, \textit{not in GEM5 documentation}.}
    \label{fig:MixedExample}
\end{figure*}

Fig.~\ref{fig:pathQuantity-all} provides further insight into the behavior of the AutoFlows++ by showing the distribution of accepted message sequences across different path lengths for the GEM5 traces. 
Compared with AutoFlows and AutoModel, AutoFlows++ exhibits a clear shift in accepted instances toward longer paths. 
In particular, medium-length paths (e.g., lengths 4–7) increase substantially. 
Moreover, AutoFlows++ almost consistently increases the number of accepted messages for higher path lengths compared to AutoFlows, indicating improved recovery of valid message flows. 
This trend suggests that the proposed method more effectively reconstructs complete flow instances rather than producing fragmented sequences. 
The improvement is primarily attributed to positional indexing and energy-based path selection, which together resolve assignment ambiguity and favor structurally consistent paths. 
As a result, the inferred model better captures true communication behavior, leading to higher acceptance ratios and improved explainability.
Fig.~\ref{fig:MixedExample} presents an example message flow mined by AutoFlows++, highlighting communication behavior that is not explicitly documented in the GEM5 documentation.

\captionsetup{skip=1pt}
\captionsetup{font=small} 

\begin{table}[!t]
\renewcommand{\arraystretch}{1.3}
\caption{Synthetic Results (The Run-Time (RT) is in seconds.)}
\resizebox{\columnwidth}{!}{%
\begin{threeparttable}
\centering
\begin{tabular}{|c||c|c|c|c|c|c|c|c|c|c|}
\hline 
\multicolumn{2}{ | c || }{}                                      & \multicolumn{3}{c |}{Large-20 (10900\tnote{*} )}         & \multicolumn{3}{c |}{Large-10 (4360\tnote{*} )}          & \multicolumn{3}{c |}{Small-20 (3680\tnote{*} )}           \\ \cline{3-11}
\multicolumn{2}{ | c || }{}                                      & RT          & Size         & Ratio             & RT          & Size          & Ratio            & RT          & Size          & Ratio             \\ 
\hline
\hline
\multicolumn{2}{ | c || }{AutoFlows++}                           & \textbf{14} & {120}        & \textbf{98.62\%}  & \textbf{10} & {126}         & \textbf{99.31\%} & \textbf{6}  & {62}          & \textbf{99.51\%}  \\ \hline
\multicolumn{2}{ | c || }{AutoFlows \cite{AutoFlows}}             & 197         & 134          & 90.26\%           & 154         & 132           & 90.47\%          & 102         & 77            & 89.57\%           \\ \hline
\multicolumn{2}{ | c || }{AutoModel \cite{OurTCAD}}              & {68}        & \textbf{103} & 57.12\%           & {50}        & \textbf{107}  & 63.69\%          & {44}        & \textbf{61}   & 69.97\%           \\ \hline
\end{tabular}
\begin{tablenotes}
\footnotesize
\item[*] message counts.
\end{tablenotes}
\end{threeparttable}
}
\label{table:syntheticResults}
\end{table}

\begin{table}[!t]
\renewcommand{\arraystretch}{1.3}
\caption{GEM5 Results (The Run-Time (RT) is in hours:minutes.)}
\resizebox{\columnwidth}{!}{%
\begin{threeparttable}
\centering
\begin{tabular}{|c|c|c|c|c|c|c|c|c|c|c|}
\hline
\multicolumn{2}{ | c | }{}                                      & \multicolumn{3}{c |}{Threads (7649395\tnote{*} )}         & \multicolumn{3}{c |}{Snoop (485497\tnote{*} )}            & \multicolumn{3}{c |}{Full-System ($8.4\times 10^{9}$\tnote{*} )} \\ \cline{3-11}
\multicolumn{2}{ | c | }{}                                      & RT            & Size         & Ratio            & RT            & Size         & Ratio            & RT            & Size         & Ratio                  \\ \hline
\multicolumn{2}{ | c | }{AutoFlows++}                           & \textbf{0:17} & \textbf{194} & \textbf{98.91\%} & \textbf{0:14} & \textbf{168} & \textbf{95.74\%} & \textbf{0:32} & \textbf{198} & \textbf{92.15\%}       \\ \hline
\multicolumn{2}{ | c | }{AutoFlows \cite{AutoFlows}}            & 2:53          & 234          & 97.92\%          & 0:42          & 191          & 92.67\%          & 9:46          & 228          & 90.04\%                \\ \hline
\multicolumn{2}{ | c | }{AutoModel \cite{OurTCAD}}              & {1:26}        & {216}        & 96.31\%          & {0:28}        & {177}        & 88.79\%          & {4:11}        & {214}        & 70.67\%                \\ \hline
\end{tabular}
\begin{tablenotes}
\footnotesize
\item[*] message counts.
\end{tablenotes}
\end{threeparttable}
}
\label{table:gem5results}
\end{table}

\subsection{Discussion}

The experimental results demonstrate that AutoFlows++ achieves consistent improvements in accuracy, model quality, and scalability across both synthetic and real-world SoC communication traces. 
These gains can be attributed to the complementary roles of hierarchical mining, energy-based path ranking, and positional indexing, which collectively address the key challenges of concurrency and ambiguity in trace-based flow inference.

First, the substantial improvement in acceptance ratio observed across all benchmarks highlights the effectiveness of decomposing the mining process into local and global stages. 
By performing interface slicing prior to global inference, AutoFlows++ eliminates a large portion of spurious causal relations that would otherwise propagate into candidate paths. 
This reduction in noise significantly stabilizes the global search space, particularly under heavy interleaving, where purely global approaches tend to produce inconsistent or incomplete flow assignments.
Second, the path energy model provides a principled mechanism for ranking candidate flows based on statistical support, structural consistency, and temporal coherence. 
Unlike heuristic scoring methods, the energy formulation enables a unified optimization perspective that naturally suppresses invalid path combinations while prioritizing flows that are well-supported by both local evidence and global structure. 
This contributes not only to improved accuracy but also to more compact and interpretable models, as reflected in the reduced model sizes across benchmarks.
Third, positional indexing plays a critical role in resolving ambiguity during model evaluation. 
In highly concurrent traces, multiple flow instances may compete for the same message occurrences, leading to incorrect assignments under sequential matching strategies. 
By incorporating positional constraints, AutoFlows++ enforces temporal consistency and ensures that message assignments align with feasible execution orderings. 
This results in a more accurate reconstruction of flow instances and significantly improves explainability.

Finally, from a scalability perspective, the proposed framework demonstrates strong performance even on extremely large traces. 
The elimination of iterative refinement cycles and the adoption of a single-pass evaluation strategy substantially reduce computational overhead. 
As shown in the full-system GEM5 experiments, AutoFlows++ processes traces containing billions of messages within practical runtime limits while maintaining high accuracy. 
This indicates that the hierarchical design not only improves accuracy but also enables efficient handling of real-world workloads.


Overall, the results demonstrate that AutoFlows++ provides a robust and scalable solution for message flow mining in complex SoC systems, effectively addressing the limitations of prior approaches in handling concurrency, ambiguity, and large-scale trace data.

\section{Practical uses of the mined message flows}
\label{subsec:usecases}

The proposed message flow mining framework can be applied to a wide range of use cases, including protocol understanding and documentation, trace debugging and failure analysis, behavioral regression analysis, communication validation and trace explainability, and verification analysis support.

\subsection{Protocol Understanding and Documentation}

Modern SoC platforms integrate numerous interacting components whose communication protocols evolve throughout the design lifecycle. 
Maintaining accurate documentation for these interactions is often difficult, particularly for legacy subsystems or third-party intellectual property blocks.

AutoFlows++ reconstructs message flows directly from execution traces, producing interpretable models that summarize observed communication behavior. 
These inferred flows act as executable documentation describing how components interact during real system execution, assisting engineers in understanding system behavior without relying solely on manually maintained specifications.

\subsection{Trace Debugging and Failure Analysis}

Debugging large-scale SoC systems frequently involves analyzing long execution traces containing heavily interleaved transactions. 
Identifying which messages belong to valid protocol executions can be challenging using manual inspection alone.

AutoFlows++ improves trace interpretability by associating messages with inferred communication flows and identifying messages that cannot be explained by valid flow instances. 
Such unexplained events may indicate protocol violations, unexpected interactions, or incomplete transaction executions. 
Consequently, the mined flow models provide useful guidance for narrowing debugging efforts and accelerating failure localization during simulation or post-silicon validation.

\subsection{Behavioral Regression Analysis}

During iterative hardware development, architectural changes, optimization passes, or software updates may unintentionally alter system communication behavior. 
Detecting such behavioral drift is difficult when relying only on raw execution traces.

Because AutoFlows++ extracts flow models representing observed communication behavior, inferred models from different design revisions can be compared directly. 
Differences in mined flows reveal changes in communication patterns, enabling automated regression analysis of system behavior across design versions.

\subsection{Communication Validation and Trace Explainability}

AutoFlows++ evaluates how well inferred flows explain observed execution traces through a position-aware evaluation procedure guided by acceptance-based metrics. 
These measures provide quantitative insight into the extent to which execution traces can be consistently decomposed into valid communication flows.

Such explainability analysis supports validation workflows by identifying portions of traces that deviate from expected behavior, helping engineers assess protocol consistency across workloads and simulation configurations.

\subsection{Support for Verification Analysis}

The communication flows inferred by AutoFlows++ provide high-level behavioral abstractions that complement existing verification methodologies. 
Instead of reasoning directly about large volumes of raw trace data, verification engineers can analyze compact flow representations summarizing exercised communication scenarios.

These abstractions can assist verification tasks such as identifying exercised interaction patterns or guiding further investigation of communication behaviors observed during simulation.

AutoFlows++ therefore serves as a practical tool for transforming raw execution traces into interpretable communication models, enabling improved understanding, debugging, and validation of complex SoC systems.

\section{Conclusions}
\label{sec:conclusion}

This paper presents AutoFlows++, a hierarchical framework for mining message flows from communication traces of complex system-on-chip (SoC) designs. 
By decomposing the mining process into local and global stages, the proposed approach effectively addresses the challenges of concurrency, interleaving, and ambiguity that arise in large-scale system executions.

At the core of AutoFlows++ are three key innovations. 
First, interface slicing enables accurate extraction of locally consistent binary relations, significantly reducing spurious dependencies prior to global inference. 
Second, the path energy model formulates candidate flow selection as an optimization problem, integrating statistical confidence, structural support, and temporal consistency into a unified ranking mechanism. 
Third, positional indexing introduces a position-aware evaluation strategy that resolves assignment ambiguity and improves the correctness and explainability of inferred flows.

Experimental results on both synthetic benchmarks and realistic architectural traces generated from GEM5 demonstrate that AutoFlows++ achieves substantial improvements in accuracy, model compactness, and runtime efficiency compared to prior approaches. 
In particular, the proposed framework consistently attains higher acceptance ratios under heavy interleaving while maintaining scalability to traces containing billions of messages.

Overall, AutoFlows++ advances the state of the art in trace-based communication mining by providing a scalable, interpretable, and robust solution for reconstructing system-level message flows. 
These capabilities position AutoFlows++ as a practical and effective solution for enabling automated specification extraction and improving the reliability of modern SoC validation workflows.

\section{Future Works}
\label{sec:future-work}

While AutoFlows++ demonstrates strong performance in accuracy, scalability, and robustness, several directions remain for future exploration. 
First, extending the framework to incorporate richer semantic information, such as message attributes or protocol-specific constraints, could further improve disambiguation in highly complex communication scenarios. 
Second, integrating lightweight learning-based techniques with the existing energy formulation may enable adaptive weighting of model components and improve generalization across diverse system architectures. 
Third, incorporating partial supervision or user feedback could guide the mining process toward domain-specific flows of interest, enhancing practical usability in industrial settings. 
Additionally, exploring online or incremental variants of the framework would enable real-time analysis of streaming traces, which is particularly relevant for post-silicon validation and runtime monitoring. 
Finally, extending the approach to jointly infer both control-flow and data-flow dependencies could provide a more comprehensive understanding of system behavior and further strengthen its applicability to large-scale hardware and software co-design environments.
\section*{Acknowledgment}

This work is supported by the National Science Foundation under Grant No. 2434247. 
Any opinions, findings, and conclusions or recommendations expressed in this material are those of the author(s) and do not necessarily reflect the views of the funding agencies.

\vspace{-60mm}
\begin{IEEEbiography}
[{\includegraphics[width=1in,height=1.25in]{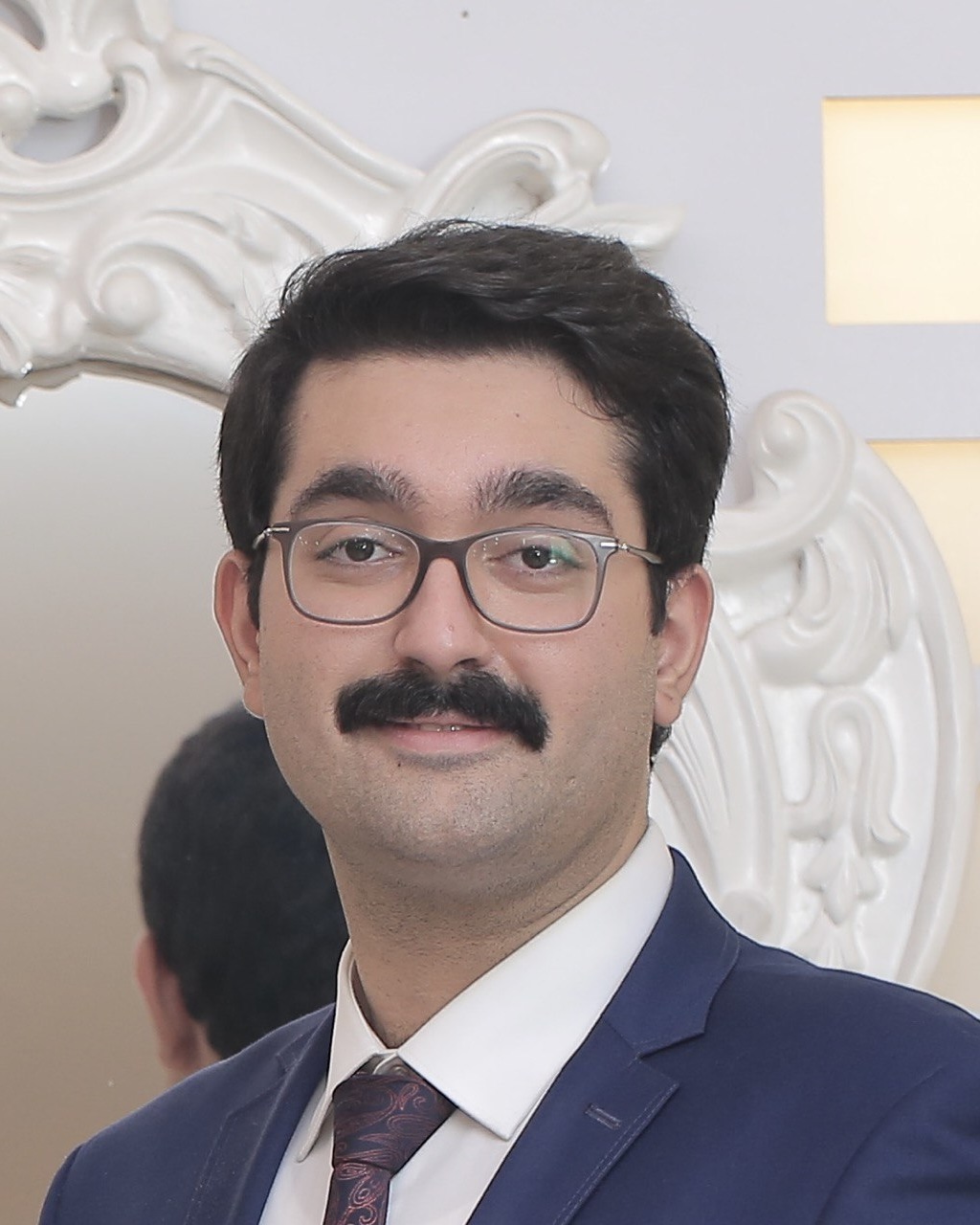}}]{Bardia Nadimi} received the B.S. degree in computer-hardware engineering and the M.S. degree in computer system architecture both from Shahid Beheshti University, Tehran, Iran, in 2017, and 2020 respectively. He is currently a Ph.D. candidate in computer architecture with the Bellini College of AI, Cybersecurity, and Computing, University of South Florida, Tampa, Florida, USA. His research interests include large language models for electronic design automation and hardware verification.
\end{IEEEbiography}
\vspace{-60mm}
\begin{IEEEbiography}
[{\includegraphics[width=1in,height=1.25in]{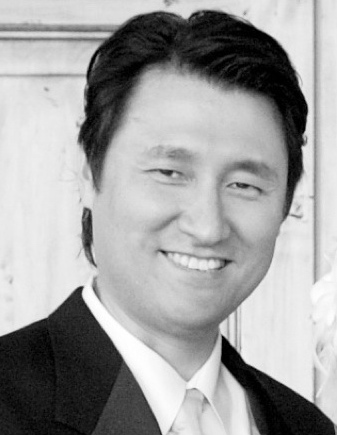}}] {Hao Zheng} received his Ph.D. in Electrical and Computer Engineering from the University of Utah in 2001. He joined the Department of Computer Science and Engineering at the University of South Florida in 2004, where he is currently an Associate Professor in the Bellini College of AI, Cybersecurity, and Computing. His research interests broadly span computer architecture, embedded systems, VLSI design, and electronic design automation (EDA). His recent work focuses on developing efficient techniques for system-on-chip (SoC) validation and debugging, runtime monitoring mechanisms for SoC security, and the application of AI to EDA. He is also interested in formal methods for verifying embedded and cyber-physical systems. Dr. Zheng is a recipient of the NSF CAREER Award (2006), the USF Outstanding Research Achievement Award (2007), and Best Paper Awards at the SPIN Symposium on Model Checking Software (2014) and the International Symposium on Quality Electronic Design (2025). He has served on program committees of numerous conferences and is currently an Associate Editor of the IEEE Transactions on Computer-Aided Design of Integrated Circuits and Systems.
\end{IEEEbiography}

\end{document}